\documentclass[aps,prb,twocolumn]{revtex4}
\usepackage{amsfonts}
\usepackage{amsmath}
\usepackage{graphicx}
\usepackage{dsfont}
\usepackage{diagbox}
\usepackage{array}
\usepackage{amssymb}
\usepackage{bbm}

\newcommand{\PreserveBackslash}[1]{\let\temp=\\#1\let\\=\temp}
\newcolumntype{C}[1]{>{\PreserveBackslash\centering}p{#1}}
\newcolumntype{R}[1]{>{\PreserveBackslash\raggedleft}p{#1}}
\newcolumntype{L}[1]{>{\PreserveBackslash\raggedright}p{#1}}

\begin{document}

\title{Enriched classification of parafermionic gapped phases with time
reversal symmetry}
\author{Wen-Tao Xu$^{1}$ and Guang-Ming Zhang$^{1,2}$}
\affiliation{$^{1}$State Key Laboratory of Low-Dimensional Quantum Physics and Department
of Physics, Tsinghua University, Beijing 100084, China. \\
$^{2}$Collaborative Innovation Center of Quantum Matter, Beijing 100084,
China.}
\date{\today}

\begin{abstract}
Based on the recently established parafermionic matrix product states, we
study the classification of one-dimensional gapped phases of parafermions
with the time reversal (TR) symmetry satisfying $T^{2}=1$. Without extra
symmetry, it has been found that $\mathbb{Z}_{p}$ parafermionic gapped
phases can be classified as topological phases, spontaneous symmetry
breaking (SSB) phases, and a trivial phase, which are uniquely labelled by
the divisors $n$ of $p$. In the presence of TR symmetry, however, the
enriched classification is characterized by three indices $n$, $\kappa $ and
$\mu $, where $\kappa \in \mathbb{Z}_{2}$ denotes the linear or projective
TR actions on the edges, and $\mu \in \mathbb{Z}_{2}$ indicates the
commutation relations between the TR and (fractionalized) charge operator.
For the $\mathbb{Z}_{r}$ symmetric parafermionic ground states, where $r=p$
for trivial or topological phases, and $r=p/n$ for SSB phases, the original
gapped phases with odd $r$ are divided into two phases, while those phases
with even $r$ are further separated into four phases. The gapped
parafermionic phases with the TR symmetry include the symmetry protected
topological phases, symmetry enriched topological phases, and the SSB
coexisting symmetry protected topological phases. From analyzing the
structures and symmetries of their reduced density matrices of those
resulting topological phases, we can obtain the topological protected
degeneracies of their entanglement spectra.
\end{abstract}

\maketitle

\section{Introduction}

Topological phases of matter and their classification have attracted
intensive interests in condensed matter physics. One of the important works
among various researches are the classification of the topological
insulators and topological superconductors\cite%
{Classification3D,AIPconf,PeriodicTable,tenfold}. However, these topological
insulators or superconductors are phases of non-interacting fermions, and
the classifications are broken down when the local interactions are included%
\cite{FidkowshiKitaev1}. It has been known that one-dimensional interacting
fermions without extra symmetry are classified as two phases, and the
classification of those with the time reversal (TR)\ symmetry are given by
the $\mathbb{Z}_{8}$ group structure\cite{FidkowshiKitaev,1Dfermion-Ent}.
These pioneer works pave the way for studying the phases of strongly
interacting fermions in higher dimensions.

Moreover, those one-dimensional fermionic topological phases possess the
Majorana edge zero modes, which have the potentials in fault-tolerant
quantum computation\cite{RMP}. In order to perform a more general quantum
computation, the exotic parafermion zero modes or fractionalized Majorana
zero modes have been proposed\cite{Quantum computing with parafermions}. The
parafermion zero modes can be generated from an effective one-dimensional
chain, and the classification of parafermion chains without extra symmetry
has been performed based on the symmetry fractionalization on the edges\cite%
{Pollmann2013,Quella}. Such a classification is beyond the framework for the
one-dimensional interacting fermion chains. We have noticed that the
classification of parafermion chains with the TR symmetry has been briefly
discussed\cite{Meidan}, but the complete classification scheme has not been
established yet.

Although the possible one-dimensional gapped phases of parafermions can be
classified, the structures of their ground state wavefunctions are still
unknown. For bosonic/spin systems, their ground state wavefunctions can be
expressed by matrix product states (MPS), all relevant information of
topological properties are encoded in the local tensors of MPS, and the
complete classification of one-dimensional bosonic/spin systems can be
implemented within the MPS formalism\cite%
{Chen-Gu-Wen-2011,Chen-Gu-Wen-2011(2),Schuch}. Recently, a framework of
fermionic MPS has been proposed, and all possible one-dimensional
topological phases of interacting fermions have been classified\cite%
{Bultinck2017,Kapustin2016}. In order to have a deeper understanding of
topological phases of parafermions, we have generalized the MPS formalism to
the parafermion systems\cite{XuZhang} and performed the classification of
gapped phases of parafermion chains\cite{XuZhang2}. Without extra symmetry,
it has been shown that the $\mathbb{Z}_{p}$ parafermionic gapped phases can
be classified as topological phases, spontaneous symmetry breaking (SSB)
phases, and a trivial phase, each phase is uniquely labelled by the divisor $%
n$ of $p$.

In this paper, using the parafermionic MPS we extend our classification
scheme to the one-dimensional gapped phases of parafermions \textit{with}
the time reversal (TR) symmetry. First of all, we carefully consider the
experimental realizations of the parafermion chains, and show that the TR
transformation on the basis of Fock space of parafermions is just to take
complex conjugation, corresponding to the BDI class with $T^{2}=1$. When the
TR symmetry is imposed on the parafermionic MPS, the possible gapped phases
are enriched and classified by three indices $n$, $\kappa $ and $\mu $,
where $\kappa \in \mathbb{Z}_{2}$ corresponds to the linear or projective
actions of TR symmetry on the edges, and $\mu \in \mathbb{Z}_{2}$ describes
the commutation relations between the TR and (fractionalized) charge
operator at the virtual degrees of freedom.
The resulting gapped parafermionic phases include the symmetry protected
topological (SPT) phases, symmetry enriched topological phases, and the SSB
coexisting SPT phases. Furthermore, we systematically analyze the structures
and symmetries of reduced density matrices for those resulting topological
phases, and derive the topological protected degeneracies of their
entanglement spectra (ES).

The paper is organized as follows. In Sec. II, the TR transformation of
parafermions is discussed according to the experimental realization setups,
and in Sec. III the classification of parafermionic MPS without extra
symmetry is briefly reviewed. In Sec. IV, the first two specific examples of
$\mathbb{Z}_{3}$ and $\mathbb{Z}_{4}$ parafermions with the TR symmetry are
considered in detail separately. The general classification of $\mathbb{Z}%
_{p}$ parafermions with TR symmetry is presented in Sec. V, and the
entanglement spectra of the topological phases are analyzed by using the
symmetries of reduced density matrices in Sec. VI. Finally, in Sec. VII we
summarize the classification results in terms of group cohomology.

\section{Time reversal symmetry for parafermions}

It is well known that the $\mathbb{Z}_{p}$ spin operators $\sigma _{l}$ and $%
\tau _{l}$ are the generalization of Pauli matrices $\sigma ^{x} $ and $%
\sigma ^{z}$. They satisfy the following relations
\begin{equation}
\sigma _{l}^{p}=\tau _{l}^{p}=1,\ \sigma _{l}\tau _{m}=\omega _{p}\tau
_{m}\sigma _{l},
\end{equation}%
where $\omega _{p}=e^{i2\pi /p}$, $\sigma ^{\dagger }=\sigma ^{p-1}$ and $%
\tau ^{\dagger }=\tau ^{p-1}$. The $\mathbb{Z}_{p}$ parafermion operators
can be introduced via the generalized Jordan-Wigner transformation\cite%
{FradkinKadonoff,AlcarazKoberle}:
\begin{equation}
\chi _{2l-1}=\left( \prod_{k<l}\tau _{k}\right) \sigma _{l},\chi
_{2l}=-e^{i\pi /p}\left( \prod_{k\leqslant l}\tau _{k}\right) \sigma _{l},
\end{equation}%
which satisfy the generalized Clifford algebra:
\begin{equation}
\chi _{l}^{p}=1,\quad \chi _{l}\chi _{m}=\omega _{p}\chi _{m}\chi _{l},\
\text{for}\ l<m.
\end{equation}%
So it is natural that the TR symmetry for parafermions has to be considered
from the $\mathbb{Z}_{p}$ spin operators\cite{F-Iemini,R-S-K-Mong}. However,
the resulting transformation is not meaningful, because the $\mathbb{Z}_{p}$
spin operators do not directly correspond to any physically realized
operators.

In order to find a well-defined TR symmetry, we consider an experimental
setup, which can realize the $\mathbb{Z}_{2m}$ parafermion modes ($m$ is an
odd integer) from a fractional topological insulator\cite%
{M-Cheng,FracMajorana,ExoticNonAbelian}. This setup can be viewed as two
copies of $\nu =\pm 1/m$ fractional quantum Hall states. The edges of this
system are gapped out in proximity to ferromagnetic or superconducting
regions, and the $\mathbb{Z}_{2m}$ parafermions live in the domain walls
between the superconducting and ferromagnetic regions, as shown in Fig. \ref%
{FM-SC}. However, this setup can not realize the $\mathbb{Z}_{p}$
parafermions with odd $p$. But there are many other proposals for the $%
\mathbb{Z}_{p}$ parafermions with odd $p$, such as bosonic or $\nu =2/3$
fractional quantum Hall states\cite{UltarColdBoson,ParafermionArray}. Here
we only focus on the parafermions realized at the edges of fractional
topological insulator, the other setups can be discussed similarly.
\begin{figure}[tbp]
\centering
\includegraphics[width=8.7cm]{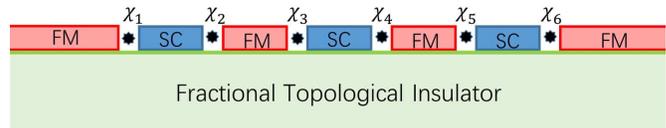}
\caption{The edges of fractional topological insulator are gapped in
proximity to the superconducting and ferromagetic regions. Parafermion zero
modes live at the domain walls.}
\label{FM-SC}
\end{figure}

The usual TR transformation for spin-$1/2$ electrons is defined by $%
Ta_{\uparrow }T^{-1}=a_{\downarrow }$, $Ta_{\downarrow }T^{-1}=-a_{\uparrow
} $ and $T^{2}=-1$, where $a_{\uparrow }$ and $a_{\downarrow }$ are the
annihilation operators of spin-up and spin-down electrons, respectively.
Since the ferromagnets in proximity to the fractional topological insulator
induce backscatterings between the two counter propagating edge modes, we
have to include the Zeeman terms ($\lambda a_{\uparrow }^{\dagger
}a_{\downarrow }+h.c.$), which explicitly break this \textit{usual} TR
symmetry. Although there is a proposal realizing the Kramers pairs of
parafermions with $T^{2}=-1$ in the absence of magnetic field\cite%
{ParafermionKramersPair}, the original setup is nevertheless invariant under
a \textit{modified} TR symmetry\cite{Meidan}
\begin{equation}
Ta_{\uparrow }T^{-1}=a_{\downarrow },\ Ta_{\downarrow }T^{-1}=a_{\uparrow },%
\text{ }T^{2}=1.
\end{equation}%
Under such a TR transformation, the electron charge remains unchanged but
the electron spin is flipped.

Based on the bosonization of the edge theory\cite%
{M-Cheng,FracMajorana,ExoticNonAbelian}, the $\mathbb{Z}_{p}$ spin operators
have the correspondences
\begin{equation}
\sigma _{l}\rightarrow e^{i\psi _{l}},\text{ }\tau _{l}\rightarrow e^{i\pi
\rho _{l}^{e}},
\end{equation}%
where $\psi _{l}$ denotes the non-chiral bosonic field whose derivative is
the electron spin density, and $\rho _{l}^{e}$ is the electron charge
density. When the TR symmetry $T^{2}=1$ is applied, it can be proven that
\begin{equation}
T\psi _{l}T^{-1}=-\psi _{l},\text{ }T\rho _{l}^{e}T^{-1}=\rho _{l}^{e}.
\end{equation}%
So the TR transformations for $\mathbb{Z}_{p}$ spin operators are obtained
as
\begin{equation}
T\sigma _{l}T^{-1}=\sigma _{l},\text{ }T\tau _{l}T^{-1}=\tau _{l}^{\dagger }.
\label{TRspin}
\end{equation}

When writing the wavefunctions for parafermions, we use the Fock space of
parafermions with the basis denoted as $|i_{1}i_{2}\cdots i_{L}\rangle $,
where $i_{l}\in \mathbb{Z}_{p}$ are the quantum numbers of $p$-dimensional
local Hilbert spaces. With the help of Fock parafermions\cite{CobaneraOrtiz}%
, it can be proved that
\begin{eqnarray}
\chi _{2l}|i_{1}\cdots i_{l}\cdots \rangle  &=&-e^{\frac{i2\pi }{p}\left(
\sum_{k\leq l}i_{k}+\frac{1}{2}\right) }|i_{1}\cdots i_{l}-1\cdots \rangle ,
\notag \\
\chi _{2l-1}|i_{1}\cdots i_{l}\cdots \rangle  &=&e^{\frac{i2\pi }{p}%
\sum_{k<l}i_{k}}|i_{1}\cdots i_{l}-1\cdots \rangle .
\end{eqnarray}%
Because we can write the $\mathbb{Z}_{p}$ spin operator in terms of the
parafermions as $\tau _{l}=-\omega _{p}^{-1/2}\chi _{2l-1}^{\dagger }\chi
_{2l}$, the basis of Fock space become the eigenstates of $\tau _{l}$,
\begin{equation}
\tau _{l}|i_{1}\cdots i_{l}\cdots i_{L}\rangle =\omega
_{p}^{i_{l}}|i_{1}\cdots i_{l}\cdots i_{L}\rangle .
\end{equation}%
According to the TR transformation of $\tau _{l}$, the basis of Fock space
are invariant under the TR transformation
\begin{equation}
T|i_{1}\cdots i_{l}\cdots i_{L}\rangle =|i_{1}\cdots i_{l}\cdots
i_{L}\rangle .
\end{equation}%
Taking into account the expression $\tau \sim e^{i\pi \rho _{e}}$, the
physical significance of the Fock basis is the electron charges of
quasi-particles modulo $p$ in the SC domains\cite{FracMajorana}, and it is
reasonable that the Fock basis keep invariant under the TR symmetry.

Actually, via the TR transformation, another set of parafermions can be
defined by
\begin{equation}
\eta _{l}=T\chi _{l}T^{-1},
\end{equation}%
which can also be expressed in terms of $\mathbb{Z}_{p}$ spin operators
\begin{equation}
\eta _{2l-1}=\left( \prod_{k<l}\tau _{k}^{\dagger }\right) \sigma _{l},\text{
}\eta _{2l}=-e^{-i\pi /p}\left( \prod_{k\leqslant l}\tau _{k}^{\dagger
}\right) \sigma _{l}.
\end{equation}%
This new type of $\mathbb{Z}_{p}$ parafermions satisfy the relations
\begin{equation}
\eta _{l}^{p}=1,\text{ }\eta _{l}\eta _{m}=\omega _{p}^{-1}\eta _{m}\eta
_{l},\text{ for }l<m.
\end{equation}%
Because the fractional topological insulator has two separate counter
propagating edge modes, these two types of parafermions $\chi _{l}$ and $%
\eta _{l}$ are the TR counterparts with each other\cite%
{FracMajorana,ParafermionArray,UniversalComputation}.

\section{Parafermionic MPS without extra symmetry}

The parafermion chains without extra symmetry other than the $\mathbb{Z}_{p}$
charge symmetry have been classified\cite{Pollmann2013,Quella}. Such a
classification can also be performed in the framework of the parafermionic
MPS\cite{XuZhang2}. In the following, we will brief review the construction
and classification of the $\mathbb{Z}_{p}$ parafermionic MPS.

For the parafermion systems, the wavefunctions or operators have the
definite charges, thus the Hilbert space is the graded vector space endowed
with the structure
\begin{equation}
\mathbb{H}=\mathbb{H}_{0}\oplus \mathbb{H}_{1}\oplus \cdots \oplus \mathbb{H}%
_{p-1},
\end{equation}%
where $\mathbb{H}_{r}$ with $r\in \mathbb{Z}_{p}$ are the charge-$r$
subspaces. In addition, the local tensors $A_{\alpha ,\beta }^{[i]}$ of MPS
with physical index $i$ and virtual indices $\alpha $ and $\beta $ also have
definite $\mathbb{Z}_{p}$ charges, determined by the three indices of the
tensors. Without loss of generality, one can use the convention that all
local tensors are charge-0. If we denote the charges of the indices $%
i,\alpha ,\beta $ with the labels $|i|,|\alpha |,|\beta |$, there exist
constrains $|i|+|\alpha |-|\beta |=0$ for all $A_{\alpha \beta }^{[i]}$,
where the charge of index $\beta $ is inverted because the opposite charge
parafermions in neighboring sites form the charge-0 bonds. Then the local
matrices as the components of the local tensors can be expressed as\cite%
{XuZhang2}
\begin{eqnarray}
A^{[i]} &=&\left[
\begin{array}{ccccc}
a_{0}^{[i]} & 0 & 0 & \cdots & 0 \\
0 & a_{1}^{[i]} & 0 & \cdots & 0 \\
0 & 0 & a_{2}^{[i]} & \cdots & 0 \\
\vdots & \vdots & \vdots & \ddots & 0 \\
0 & 0 & 0 & 0 & a_{p-1}^{[i]}%
\end{array}%
\right] ,|i|=0,  \notag  \label{standrad form} \\
A^{[i]} &=&\left[
\begin{array}{ccccc}
0 & a_{0}^{[i]} & 0 & \cdots & 0 \\
0 & 0 & a_{1}^{[i]} & \cdots & 0 \\
0 & 0 & 0 & \cdots & 0 \\
\vdots & \vdots & \vdots & \ddots & \vdots \\
a_{p-1}^{[i]} & 0 & 0 & 0 & 0%
\end{array}%
\right] ,|i|=1,  \notag \\
&\vdots &\quad  \notag \\
A^{[i]} &=&\left[
\begin{array}{ccccc}
0 & 0 & 0 & \cdots & a_{0}^{[i]} \\
a_{1}^{[i]} & 0 & 0 & \cdots & 0 \\
0 & a_{2}^{[i]} & 0 & \cdots & 0 \\
\vdots & \vdots & \vdots & \ddots & \vdots \\
0 & 0 & 0 & a_{p-1}^{[i]} & 0%
\end{array}%
\right] ,|i|=p-1,
\end{eqnarray}%
where $a_{r}^{[i]}$ with $r\in \mathbb{Z}_{p}$ are the sub-block matrices.

For the convenience of discussion, we introduce two $n\times n$ block
matrices
\begin{eqnarray}
\mathcal{Y}_{n} &=&\left[
\begin{array}{ccccc}
0 & \mathbbm{1} & 0 & \cdots & 0 \\
0 & 0 & \mathbbm{1} & \cdots & 0 \\
0 & 0 & 0 & \cdots & 0 \\
\vdots & \vdots & \vdots & \ddots & \mathbbm{1} \\
\mathbbm{1} & 0 & 0 & 0 & 0%
\end{array}%
\right] ,  \notag \\
\mathcal{Q}_{n} &=&\text{diag}(\mathbbm{1},\omega \mathbbm{1},\omega ^{2}%
\mathbbm{1},\cdots ,\omega ^{n-1}\mathbbm{1}),
\end{eqnarray}%
where the dimensions of the sub-block identities $\mathbbm{1}$ should be
coincided with that of the sub-block matrices $a_{r}^{[i]}$. The matrix $%
\mathcal{Q}_{n}$ measures the $\mathbb{Z}_{n}$ charge and the matrix $%
\mathcal{Y}_{n}$ flips the $\mathbb{Z}_{n}$ charge. When the identity
matrices $\mathbbm{1}$ are reduced to the number $1$, we will denote $%
\mathcal{Y}_{n}$ by $Y_{n}$ and $\mathcal{Q}_{n}$ by $Q_{n}$. With these
definitions, Eq. (\ref{standrad form}) can be written into a more concise
form:
\begin{equation}
A^{[i]}=\text{diag}\left( a_{0}^{[i]},a_{1}^{[i]},\cdots
,a_{p-1}^{[i]}\right) \times \mathcal{Y}_{p}^{|i|}.  \label{trivial}
\end{equation}

Supposing that all $a_{r}^{[i]}$ can not become equal under any gauge
transformations, the MPS generated by the matrices of Eq.(\ref{trivial})
belong to the trivial phase. If all sub-blocks can be equal under a gauge
transformation, i.e., $a_{r}^{[i]}=a^{[i]}$, the local matrices of MPS can
be written as
\begin{equation}
A^{[i]}=Y_{p}^{|i|}\otimes a^{[i]}.  \label{non-trivial}
\end{equation}%
The generated MPS represent the ground states of a topological phase with
unpaired $\mathbb{Z}_{p}$ parafermion zero edge modes\cite{XuZhang2}.
Moreover, if $a_{r}^{[i]}=a_{(r+p/n)\bmod p}^{[i]}$ under a gauge
transformation, where $n$ is a divisor of $p$, there are $p/n$ unequal
sub-block matrices $a_{r}^{[i]}$ with $r\in \mathbb{Z}_{p/n}$, and there are
two different situations. If $n$ and $p/n$ are mutually prime, using a
charge-preserving gauge transformation represented by a permutation matrix,
the local matrices can be transformed into
\begin{eqnarray}
A^{[i]} &=&Y_{n}^{|i|}\otimes d^{[i]},  \notag  \label{non-trivial2} \\
d^{[i]} &=&\text{diag}\left( a_{0}^{[i]},\cdots ,a_{p/n-1}^{[i]}\right)
\times \mathcal{Y}_{p/n}^{|i|}.
\end{eqnarray}%
The generated MPS correspond to a $\mathbb{Z}_{p}$ symmetric topological
phase with unpaired $\mathbb{Z}_{n}$ parafermion zero edge modes\cite%
{XuZhang2}. Such a topological phase is characterized by the $n$-fold
degenerate ES.

In the case that $n$ and $p/n$ are not mutually prime, it is impossible that
the local matrices can be gauge transformed into the form of Eq. (\ref%
{non-trivial2}), because $Y_p\sim Y_n\otimes Y_{p/n}$ only if $n$ and $p/n$
are mutually prime. With $\tilde{Q}_{n}=\text{diag}\left( 1,\omega
_{p}^{1},\omega _{p}^{2},\cdots ,\omega _{p}^{n-1}\right) $, we can write $%
Y_{p}\sim \tilde{Q}_{n}\otimes Y_{p/n}$, and then the local matrices can be
gauge transformed into
\begin{equation}
A^{[i]}=\tilde{Q}_{n}^{|i|}\otimes d^{[i]},  \label{SSB standrad form}
\end{equation}%
where the gauge transformation breaks the $\mathbb{Z}_{p}$ charge symmetry
but preserves the $\mathbb{Z}_{p/n}$ charge symmetry. The block diagonal
forms of $A^{[i]}$ represent a SSB phase, where the $\mathbb{Z}_{p}$
symmetry is spontaneously broken down to $\mathbb{Z}_{p/n}$ symmetry\cite%
{XuZhang2}.

So the number of gapped phases of $\mathbb{Z}_{p}$ parafermions is the same
as the number of divisors of $p$, and each divisor $n$ uniquely labels a
gapped phase. If $n$ and $p/n$ are mutually prime, they are the topological
phases with $\mathbb{Z}_{n}$ parafermion zero edge modes except the $n=1$
trivial phase. Otherwise, they are SSB phases.

\section{$\mathbb{Z}_{3}$ and $\mathbb{Z}_{4}$ parafermion phases enriched
by time reversal symmetry}

\subsection{$\mathbb{Z}_{3}$ parafermionic phases}

In Sec. II, we have shown that the basis of Fock space are invariant under
TR transformation. The matrix form of TR symmetry operator can be expressed
as $T_{ij}=\delta _{ij}K$, where $K$ is the complex conjugation operator.
Because a TR symmetric MPS are invariant under TR transformation up to a
gauge transformation $\mathcal{T}$, the projective representation of $T$,
the MPS local matrices satisfy
\begin{equation}
\sum_{j}\delta _{ij}KA^{[j]}=\bar{A}^{[i]}=\mathcal{T}^{-1}A^{[i]}\mathcal{T}%
,  \label{TR}
\end{equation}%
where $\bar{A}^{[i]}$ are the complex conjugation of $A^{[i]}$. Since $T^{2}=%
\mathbbm{1}$, we return back to the original matrices via twice TR
transformation
\begin{equation}
A^{[i]}=\mathcal{T\bar{T}}A^{[i]}\mathcal{\bar{T}}^{-1}\mathcal{T}^{-1}.
\label{TRtwice}
\end{equation}%
Because there are two distinct gapped phases for $\mathbb{Z}_{3}$
parafermion chains, their enriched classification with TR symmetry should be
discussed separately.

For the original trivial phase, the irreducible local matrices $A^{[i]}$ are
injective\cite{XuZhang2}, so the only way to fulfill Eq, (\ref{TRtwice}) is
\begin{equation}
\mathcal{T\bar{T}=}\alpha _{0}\mathbbm{1}.  \label{TTbar}
\end{equation}%
Without loss of generality, one can assume $\alpha _{0}=(-1)^{\kappa }$,
where $\kappa \in \mathbb{Z}_{2}$ labels two distinct classes of projective
representations. In addition, the systems have the intrinsic $\mathbb{Z}_{3}$
charge symmetry, the MPS are invariant under the action of the $\mathbb{Z}%
_{3}$ charge operator up to the a gauge transformation $\mathcal{Q}_{3}$:
\begin{equation}
\mathcal{Q}_{3}^{-1}A^{[i]}\mathcal{Q}_{3}=\sum_{j}\left( Q_{3}\right)
_{ij}A^{[j]}=\omega _{3}^{|i|}A^{i}.
\end{equation}%
Accordingly the MPS should also be invariant under the combined action of $T$
and $Q_{3}$, which are not commute each other. Comparing the transformations
of $T$ and $Q$ with different orders, we find that the injectivity of $%
A^{[i]}$ requires $\mathcal{T}$ with a definite $\mathbb{Z}_{3}$ charge,
i.e.,
\begin{equation}
\mathcal{Q}_{3}^{-1}\mathcal{T}\mathcal{Q}_{3}=\omega _{3}^{|\mathcal{T}|}%
\mathcal{T},\quad |\mathcal{T}|=0,1,2.  \label{charge of T}
\end{equation}%
Obviously $\mathcal{T}$ and $\mathcal{\bar{T}}$ have the same charge, and
Eq. (\ref{TTbar}) uniquely determines that $\mathcal{T}$ has charge-zero,
i.e., $|\mathcal{T}|=0$ and $\mathcal{Q}_{3}\mathcal{T}=\mathcal{T}\mathcal{Q%
}_{3}$. Unlike the fermionic MPS, the commutation relation between $\mathcal{%
Q}_{3}$ and $\mathcal{T}$ does not lead to a topological invariant. By
imposing the TR symmetry, the trivial phase of $\mathbb{Z}_{3}$ parafermions
is thus split into two TR symmetric phases labelled by $\kappa =0,1$. They
are the trivial phase ($\kappa =0$) and SPT phase ($\kappa =1$) with Kramers
doublets at the edges.

On the other hand, the irreducible $\mathbb{Z}_{3}$ parafermionic MPS for
the topological phase do not have the injective property. Since the
structures of local matrices $A^{[i]}=Y_{3}^{|i|}\otimes a^{[i]}$ are
featured by the matrix $Y_{3}$, the regular representation of the $\mathbb{Z}%
_{3}$ generator, the MPS of the topological phase are $\mathbb{Z}_{3}$
injective. Considering the special structures of $A^{[i]}$, we can without
loss of generality assume that $\mathcal{T}$ has a well-defined charge\cite%
{Bultinck2017}, i.e., one of the three matrices $\mathcal{T}_{0},\mathcal{T}%
_{1},\mathcal{T}_{2}$ with charge-$0$, charge-$1$ and charge-$2$ fulfills
the TR transformation. Because $\mathcal{Y}_{3}$ commutes with all $A^{[i]}$%
, $\mathcal{T}_{0}$, $\mathcal{T}_{1}$ and $\mathcal{T}_{2}$ are not
independent, they are connected by $\mathcal{Y}_{3}$:
\begin{equation}
\mathcal{T}_{1}=\mathcal{T}_{0}\mathcal{Y}_{3},\quad \mathcal{T}_{2}=%
\mathcal{T}_{0}\mathcal{Y}_{3}^{2}.  \label{different-T}
\end{equation}%
Moreover, the possible way satisfying Eq. (\ref{TRtwice}) is given by
\begin{equation}
\mathcal{T}_{0}\mathcal{\bar{T}}_{0}\mathcal{=}\alpha _{0}\mathbbm{1},\quad
\mathcal{T}_{1}\mathcal{\bar{T}}_{1}\mathcal{=}\alpha _{1}\mathcal{Y}%
_{3}^{2},\quad \mathcal{T}_{2}\mathcal{\bar{T}}_{2}\mathcal{=}\alpha _{2}%
\mathcal{Y}_{3}.  \label{T-square}
\end{equation}%
Analogy to the trivial phase, we generally have $\alpha _{0}=(-1)^{\kappa }$
with $\kappa =0,1$. From Eq.(\ref{different-T}) and Eq.(\ref{T-square}), we
can determine $\alpha _{1}^{3}=\alpha _{2}^{3}=\alpha _{0}$ as well as $%
\alpha _{1}=\alpha _{0}\alpha _{2}^{2}$, and another relation between $%
\alpha _{0}$ and $\alpha _{1}$ can be obtained by the commutation relation
between $\mathcal{T}_{0}$ and $\mathcal{Y}_{3}$:
\begin{equation}
\mathcal{Y}_{3}\mathcal{T}_{0}=\omega _{3}^{\mu }\mathcal{T}_{0}\mathcal{Y}%
_{3},\quad \omega _{3}^{\mu }=\alpha _{0}\alpha _{1},
\end{equation}%
where $\mu =0,1,2$. In the classification of $\mathbb{Z}_{2}$ fermion chains%
\cite{1Dfermion-Ent,Bultinck2017,FidkowshiKitaev}, $\mu $ distinguish
different phases. So we might expect that different values of $\mu $ also
label different parafermion phases, but there are some redundancies.

In general, the MPS of topological phase intrinsically have $\mathbb{Z}_{p}$
charge symmetry, which is implemented by a gauge transformation $\mathcal{Q}%
_{p}^{r}$ as $A^{\prime \lbrack i]}=\mathcal{Q}_{p}^{-r}A^{[i]}\mathcal{Q}%
_{p}^{r}=\omega ^{r|i|}A^{[i]}$, where $r\in \mathbb{Z}_{p}$, and the ground
state wavefunctions generated by $A^{[i]}$ and $A^{\prime \lbrack i]}$ are
the same. We then denote that $\mathcal{T}_{0}^{\prime }$ is the projective
charge-0 representation of $T$ associating with the local matrices $%
A^{\prime \lbrack i]}$, i.e., $\bar{A}^{\prime \lbrack i]}=\mathcal{T}%
_{0}^{\prime -1}A^{\prime \lbrack i]}\mathcal{T}_{0}^{\prime }$. Compare to
Eq.(\ref{TR}), the $\mathbb{Z}_{3}$-injectivity of $A^{[i]}$ gives rise to
\begin{equation}
\mathcal{T}_{0}^{\prime }=\mathcal{Q}_{p}^{-2r}\mathcal{T}_{0},\quad
\mathcal{Y}_{p}\mathcal{T}_{0}^{\prime }=\omega _{p}^{\mu -2r}\mathcal{T}%
_{0}^{\prime }\mathcal{Y}_{p},  \label{equivlence}
\end{equation}%
indicating that $\mu $ and $(\mu -2r)\bmod p$ should characterize the same
phase. Therefore, for even $p$, only the parity of $\mu $ can distinguish
different topological phases, while for odd $p$, all phases with different $%
\mu $ are equivalent.

As a result, only two non-trivial topological phases with the TR symmetry
labelled by $\kappa =0,1$ are obtained from the topological phase of $%
\mathbb{Z}_{3}$ parafermions, and they are referred to as the symmetry
enriched topological phases. By including the one SPT phase and the trivial
phase, there exist four phases labelled by the indices $n$ and $\kappa $,
which are summarized in the Tab. \ref{Z3}.
\begin{table}[tbp]
\caption{The classification of gapped phases of $\mathbb{Z}_{3}$
parafermions with TR symmetry. Different phases are labelled by the indices $%
n$ and $\protect\kappa $, where $\protect\kappa $ describes whether there
exist Kramers doublets at the edges.}
\label{Z3}%
\begin{tabular}{|C{1.6cm}|C{1.4cm}|C{1.4cm}|C{1.4cm}|C{1.4cm}|}
\hline Phase&\multicolumn{2}{c|}{Trivial} & \multicolumn{2}{c|}{Non-trivial} \\\hline
$n$&\multicolumn{2}{c|}{$1$} & \multicolumn{2}{c|}{$3$} \\ \hline
$\kappa$&$0$ &$1$& $0$ & $1$ \\ \hline
\end{tabular}
\end{table}

\subsection{$\mathbb{Z}_{4}$ parafermionic phases}

For the trivial phase of $\mathbb{Z}_{4}$ parafermions, we still have
\begin{equation}
\mathcal{T\bar{T}}=(-1)^{\kappa }\mathbbm{1},
\end{equation}%
where $\kappa =0$ and $1$ characterize two different gapped phases. Similar
to Eq.(\ref{charge of T}), $\mathcal{Q}_{4}^{-1}\mathcal{T}\mathcal{Q}%
_{4}=\omega _{4}^{|\mathcal{T}|}\mathcal{T}$ with $|\mathcal{T}|=0,1,2,3$
for the $\mathbb{Z}_{4}$ parafermions. $\mathcal{T}$ is further required to
have a definite $\mathbb{Z}_{4}$ charge. Since $\mathcal{T}$ and $\mathcal{%
\bar{T}}$ have the same charge, it is only possible that $\mathcal{T}$ has
charge-$0$ or charge-$2$, determined by
\begin{equation}
\mathcal{Q}_{4}\mathcal{T}=(-1)^{\mu }\mathcal{T}\mathcal{Q}_{4}
\end{equation}%
with $\mu =0,1$. Unlike the $\mathbb{Z}_{3}$ case, the charge of $\mathcal{T}
$ can take two different values and $\mu $ is thus a topological invariant.
So from the trivial phase of $\mathbb{Z}_{4}$ parafermions, there emerge
four different gapped phases labeled by $\kappa =0,1$ and $\mu =0,1$. Among
them, there are three SPT phases with the TR symmetry.

According to the structures of local matrices $A^{[i]}=Y_{4}^{|i|}\otimes
a^{[i]}$, the MPS for the non-trivial topological phase of $\mathbb{Z}_{4}$
parafermions are $\mathbb{Z}_{4}$-injective. Similar to the $\mathbb{Z}_{3}$
case, the projective representation $\mathcal{T}$ can be restricted to the
charge-$q$ matrices $\mathcal{T}_{q}$ satisfying $\mathcal{T}_{q}\mathcal{%
\bar{T}}_{q}\mathcal{=}\alpha _{q}\mathcal{Y}^{2q}$ with $q=0,1,2,3$ . Among
them, $\mathcal{T}_{0}$ is used to define two different topological
invariants like the $\mathbb{Z}_{3}$ classification:
\begin{equation}
\mathcal{T}_{0}\bar{\mathcal{T}}_{0}=(-1)^{\kappa }\mathbbm{1},\text{ }%
\mathcal{Y}_{4}\mathcal{T}_{0}=\omega ^{\mu }\mathcal{T}_{0}\mathcal{Y}_{4},
\end{equation}%
where $\kappa =0,1$ and $\mu =0,1,2,3$. From the Eq. (\ref{equivlence}), $%
\mu $ and $(\mu -2r)\bmod4$ label the same phases. Therefore, there exist
four different symmetry enriched topological phases labelled by $\kappa =0,1$
and $\mu =0,1$. $\kappa =1$ implies the existence of the Kramers degeneracy,
and $\mu $ classifies the actions of the TR transformation on parafermion
zero edge modes.

In addition, for the $\mathbb{Z}_{4}$ parafermions, there also exists a SSB
phase. According to Eq. (\ref{SSB standrad form}), the local matrices can be
expressed as
\begin{eqnarray}
A^{[i]} &=&\left[
\begin{array}{cc}
d^{[i]} & 0 \\
0 & \omega _{4}^{|i|}d^{[i]}%
\end{array}%
\right] ,  \notag \\
d^{[i]} &=&\left[
\begin{array}{cc}
a_{0}^{[i]} & 0 \\
0 & a_{1}^{[i]}%
\end{array}%
\right] ,|i|=0,2,  \notag \\
d^{[i]} &=&\left[
\begin{array}{cc}
0 & a_{0}^{[i]} \\
a_{1}^{[i]} & 0%
\end{array}%
\right] ,|i|=1,3.
\end{eqnarray}%
To fulfill the unity requirement of twice TR transformation, the projective
TR representation $\mathcal{T}$ should have the following block diagonal
form
\begin{equation}
\mathcal{T}=\left[
\begin{array}{cc}
\mathcal{T}_{0,0} & 0 \\
0 & \mathcal{T}_{1,1}%
\end{array}%
\right] .
\end{equation}%
And the TR transformations for the sub-blocks $d^{[i]}$ and $\omega
_{4}^{|i|}d^{[i]}$ yield
\begin{equation}
\mathcal{T}_{0,0}\mathcal{\bar{T}}_{0,0}=(-1)^{\kappa }\mathbbm{1},\quad
\mathcal{T}_{1,1}\mathcal{\bar{T}}_{1,1}=\alpha _{1}\mathbbm{1}.
\end{equation}%
Actually, $\mathcal{T}_{0,0}$ and $\mathcal{T}_{1,1}$ are not independent.
Considering that the MPS generated by $d^{[i]}$ are injective and $\mathbb{Z}%
_{2}$ parity symmetric, i.e., $\mathcal{Q}_{2}d^{[i]}\mathcal{Q}_{2}=\omega
_{2}^{|i|}d^{[i]}$, we then have
\begin{equation}
\mathcal{T}_{1,1}=\mathcal{Q}_{2}\mathcal{T}_{0,0},\quad \mathcal{T}_{0,0}%
\mathcal{Q}_{2}=(-1)^{\mu }\mathcal{Q}_{2}\mathcal{T}_{0,0}.
\end{equation}%
The relation between $\mathcal{T}_{0,0}$ and $\mathcal{T}_{1,1}$ leads to $%
\alpha _{1}=(-1)^{\mu +\kappa }$, and $\mu $ just indicates the $\mathbb{Z}%
_{2}$ parity of $\mathcal{T}_{0,0}$. So we can obtain four different SSB
coexisting SPT phases, labelled by $\kappa =0,1$ and $\mu =0,1$.

By including the SPT phases split from the trivial phase and four symmetry
enriched topological phases, we have obtained three different families of $%
\mathbb{Z}_{4}$ parafermionic gapped phases, each of them consists of four
different phases labelled by the indices $\kappa$ and $\mu$. The results are
summarized in the Tab. \ref{Z4}
\begin{table}[tbp]
\caption{The gapped phases of $\mathbb{Z}_{4}$ parafermions with TR symmetry
are classified by three topological indices $n$, $\protect\kappa $, $\protect%
\mu $. $\protect\kappa $ denotes the presence/absence of the Kramers
doublets at the edges of the chains. For the trivial and SSB phases, $%
\protect\mu $ describes whether the TR changes the charges of the edge
states. For the topological phase, $\protect\mu $ classifies the TR actions
on parafermion zero modes.}
\label{Z4}%
\begin{tabular}{|C{1.3cm}|C{0.4cm}|C{0.4cm}|C{0.4cm}|C{0.4cm}|C{0.5cm}|C{0.5cm}|C{0.5cm}|C{0.5cm}|C{0.4cm}|C{0.4cm}|C{0.4cm}|C{0.4cm}|}
\hline
phase & \multicolumn{4}{c|}{Trivial} & \multicolumn{4}{c|}{SSB
} & \multicolumn{4}{c|}{Non-trivial} \\ \hline
$n$ & \multicolumn{4}{c|}{1} & \multicolumn{4}{c|}{2} &
\multicolumn{4}{c|}{4} \\ \hline
$\kappa$ & 0 & 0 & 1 & 1 & 0 & 0 & 1 & 1 & 0 & 0 & 1 & 1 \\ \hline
$\mu $ & 0 & 1 & 0 & 1 & 0 & 1 & 0 & 1 & 0 & 1 & 0 & 1 \\ \hline
\end{tabular}
\end{table}

\section{Enriched classification of $\mathbb{Z}_{p}$ parafermionic phases
with TR symmetry}

The general cases are more complicated than $\mathbb{Z}_{3}$ and $\mathbb{Z}%
_{4}$ cases, because there exist various topological phases and SSB phases.
We will classify all these phases with TR symmetry in the following, and the
obtained results are summarized in Tab. \ref{GeneralCaseDeg}.
\begin{table*}[tbp]
\caption{The enriched classification of all gapped $\mathbb{Z}_{p}$
parafermion phases with the TR symmetry. For odd $p$, the original phases
always split into 2 phases, while for even $p$ the phases usually split into
$4$ phases except the SSB phases with odd $p/n$. The last two columns do not
exist for odd $p/n$. The last row shows the topological protected
degeneracy, where \textquotedblleft gcd\textquotedblright\ denotes the
greatest common divisor and $\text{gcd}(p/n,2)n$ means the degeneracy is $n$
for odd $p/n$ and $2n$ for even $p/n$. }
\label{GeneralCaseDeg}%
\begin{tabular}{|C{1.3cm}|C{0.4cm}|C{0.4cm}|C{0.65cm}|C{0.65cm}|C{0.4cm}|C{0.4cm}|C{0.4cm}|C{0.4cm}|C{0.4cm}|C{0.4cm}|C{0.4cm}|C{0.4cm}
|C{1.7cm}|C{0.4cm}|C{0.4cm}|C{0.4cm}|C{1.3cm}|C{1.3cm}|}
\hline
$p$ & \multicolumn{6}{c|}{odd} & \multicolumn{12}{c|}{even} \\ \hline
phase & \multicolumn{2}{c|}{trivial} & \multicolumn{2}{c|}{non-trivial} & \multicolumn{2}{c|}{SSB} & \multicolumn{4}{c|}{trivial} &\multicolumn{4}{c|}{non-trivial}& \multicolumn{4}{c|}{SSB} \\ \hline
divisor $n$&\multicolumn{2}{c|}{$1$} & \multicolumn{2}{c|}{$n$} & \multicolumn{2}{c|}{$n$} & \multicolumn{4}{c|}{$1$}&
\multicolumn{4}{c|}{$n$} & \multicolumn{4}{c|}{$n$} \\ \hline
$\mu $ & $0$ & $0$ & $0$ & $0$ & $0$ & $0$  & \multicolumn{2}{c|}{$0$} &
\multicolumn{2}{c|}{$1$} & \multicolumn{2}{c|}{$0$} &
\multicolumn{2}{c|}{$1$}& \multicolumn{2}{c|}{$0$} &
\multicolumn{2}{c|}{$1$(only for even $p/n$)} \\ \hline
$\kappa $& $0$ & $1$ & $0$ & $1$ & $0$ & $1$ & $0$ & $1$ & $0$ & $1$ & $0$ & $1
$ & $0$ & $1$  &$0$& $1$ & $0$ & $1$ \\ \hline
top. deg. & $1$ & $2$ & $n$ & $2n$ & $1$ & $2$ & $1$ & $2$ & $2$ & $2$  & $n$ & $2n$ & $\text{gcd}(p/n, 2)n$ &$2n$  &$1$& $2$ & $2$&$2$ \\ \hline
\end{tabular}
\end{table*}

\subsection{From the trivial phase}

The local matrices generating the MPS for the trivial phase of $\mathbb{Z}%
_{p}$ parafermions have been given by Eq. (\ref{trivial}). The injectivity
of irreducible local matrices $A^{[i]}$ and Eq. (\ref{TRtwice}) determine
\begin{equation}
\mathcal{T}\bar{\mathcal{T}}=(-1)^{\kappa }\mathbbm{1},\quad \kappa=0,1.
\end{equation}%
Eq. (\ref{charge of T}) is still valid even when generalizing to $\mathbb{Z}%
_{p}$ parafermions,
\begin{equation}  \label{ChargeOfT}
\mathcal{Q}_{p}^{-1}\mathcal{T}\mathcal{Q}_{p}=\omega _{p}^{|\mathcal{T}|}%
\mathcal{T},\quad |\mathcal{T}|\in \mathbb{Z}_{p},
\end{equation}%
which enforces that $\mathcal{T}$ has a definite $\mathbb{Z}_{p}$ charge.
Since $\mathcal{T}$ and $\mathcal{\bar{T}}$ have the same charge, $\mathcal{T%
}$ must be charge-0 for odd $p$, and charge-$0$ or charge-$p/2$ for even $p$%
, corresponding to $\mathcal{T}\mathcal{Q}_{p}=(-1)^{\mu }\mathcal{Q}_{p}%
\mathcal{T}$ with $\mu =0,1$, respectively. So the trivial phase of $\mathbb{%
Z}_{p}$ parafermions with odd $p$ is split into two different phases
labelled by $\kappa =0,1$, while the trivial phase with even $p$ is split
into four phases labelled by $\kappa =0,1$ and $\mu =0,1$. $\kappa =1$
signifies that there is the Kramers degeneracy at the each end of the
chains, while $\mu =1$ indicates that $\mathcal{T}$ shifts the $\mathbb{Z}%
_{p}$ charges of the edge states by $p/2$.

\subsection{From the non-trivial phases}

There might be many non-trivial topological phases, and each phase is
labelled by $n$ $(n\neq 1)$, satisfying $n$ and $p/n$ are mutually prime.
The local matrices of the non-trivial phases have the structures $%
A^{[i]}=Y_{n}^{|i|}\otimes d^{[i]}$, where $d^{[i]}$ can generate injective
MPS. Since $\mathcal{Y}_{n}^{|i|}$ commute with all $A^{[i]}$ which are $%
\mathbb{Z}_{n}$ injective, we can restrict $\mathcal{T}$ to the matrices $%
\mathcal{T}_{r}$ with arbitrary $\mathbb{Z}_{r}$ charges $r\in \mathbb{Z}%
_{n} $. They satisfy $\mathcal{T}_{r}\mathcal{\bar{T}}_{r}=\alpha _{r}%
\mathcal{Y}_{n}^{2r}$ and are transformed into each other by multiplying $%
\mathcal{Y}_{n}$ several times. Analogue to the previous examples, we have
\begin{equation}
\mathcal{T}_{0}\bar{\mathcal{T}}_{0}=(-1)^{\kappa }\mathbbm{1},
\end{equation}%
and the relations among $\alpha _{r}$ are determined by
\begin{equation}
\mathcal{Y}_{n}\mathcal{T}_{0}=\omega _{n}^{\mu ^{\prime }}\mathcal{T}_{0}%
\mathcal{Y}_{n},  \label{41}
\end{equation}%
where $\mu ^{\prime }$ and $(\mu ^{\prime }-2r)\bmod n$ represent the same
phases as explained before. Therefore, if $n$ is odd, different $\mu
^{\prime }$ are equivalent. But if $n$ is even, the parities of $\mu
^{\prime }$ define the equivalent classes.

It should be noticed that the MPS generated by $d^{[i]}$ have the $\mathbb{Z}%
_{p/n}$ symmetry. Eq.(\ref{41}) implies that $\mathcal{T}_{0}$ has the
structure $\mathcal{T}_{0}=Q_{n}^{\mu ^{\prime }}\otimes \mathcal{T}_{0,0}$.
Similar to Eq. (\ref{charge of T}), $\mathcal{T}_{0,0}$ should have a
definite $\mathbb{Z}_{p/n}$ charge,
\begin{equation}
\mathcal{Q}_{p/n}^{-1}\mathcal{T}_{0,0}\mathcal{Q}_{p/n}=\omega _{p/n}^{|%
\mathcal{T}_{0,0}|}\mathcal{T}_{0,0}.  \label{mu}
\end{equation}%
Then for odd $p/n$, we have $|\mathcal{T}_{0,0}|=0$, while for even $p/n$,
we have $|\mathcal{T}_{0,0}|=0$ or $p/(2n)$, corresponding to $\mathcal{T}%
_{0,p/n}\mathcal{Q}_{p/n}=(-1)^{\mu }\mathcal{Q}_{p/n}\mathcal{T}_{0,p/n}$
with topological invariant $\mu =0,1$, respectively.

Now we have three topological indices $\kappa =0,1$, $\mu ^{\prime }=0,1$
and $\mu =0,1$. The MPS matrices generally have both a trivial part and a
non-trivial part, and $\mu $ comes from the trivial part and $\mu ^{\prime }$
originates from the non-trivial part. Since they are mutually prime for the
non-trivial phases, $n$ and $n/p$ can not be even number simultaneously.
When $p$ is odd, both $\mu $ and $\mu ^{\prime }$ are equal to zero. If $p$
is even and odd $p/n$, $\mu ^{\prime }=0,1$ and $\mu =0$; while for both
even $p$ and $p/n$, $\mu =0,1$ and $\mu ^{\prime }=0$. So in Tab. \ref%
{GeneralCaseDeg} and the following discussion, we do not distinguish $\mu $
from $\mu ^{\prime }$ and both are denoted as $\mu $. When we count the
degeneracy of ES, however, we must remember which part of $\mu $ stems from.
Therefore, for odd $p$, there are two phases labelled by $\kappa =0,1$,
while for even $p$ there are four phases simply labelled by $\kappa =0,1$
and $\mu =0,1$.

\subsection{From the SSB phases}

We now consider the SSB phases, where $\mathbb{Z}_{p}$ symmetry is broken
down to $\mathbb{Z}_{p/n}$ and $n$ and $p/n$ are not mutually prime in this
case. Applying the $\mathbb{Z}_{p}$ charge matrix to all physical degrees of
freedom will shift one ground state to another degenerate ground state, and
it will go back to the original ground state after the $\mathbb{Z}_{p}$
charge operator acts $n$ times. Therefore, the local matrices can be
generally written as Eq. (\ref{SSB standrad form}), where the sub-blocks $%
\omega _{p}^{|i|r}d^{[i]}$ generate a $\mathbb{Z}_{p/n}$ symmetric ground
state, which is the same as the MPS of $\mathbb{Z}_{p/n}$ trivial phase.

Suppose that there is a short-range correlated ground state which is
invariant under only a subgroup of the whole symmetry group. It is possible
to have the gapped phases in which long-range order and SPT order coexist%
\cite{Chen-Gu-Wen-2011(2),Schuch}. To classify these phases, we can combine
the symmetry breaking and symmetry fractionalization using two sets of data:
the subgroup and the SPT order under the subgroup. Considering Eq. (\ref{TR}%
) and Eq. (\ref{TRtwice}) together, $\mathcal{T}$ should have the form
\begin{equation}
\mathcal{T}=\text{diag}(\mathcal{T}_{0,0},\mathcal{T}_{1,1},\cdots ,\mathcal{%
T}_{n-1,n-1})\times \left( P\otimes \mathbbm{1}\right) ,
\end{equation}%
where $P$ is the $n\times n$ permutation matrix occurring in classification
for SSB phases\cite{Chen-Gu-Wen-2011(2),Schuch}. When the TR symmetry is
imposed, we can in general assume that the matrices of the zeroth ground
state satisfy $\bar{d}^{[i]}=\mathcal{T}_{0,0}^{-1}d^{[i]}\mathcal{T}_{0,0}$%
. However, the matrices $\omega _{p}^{r|i|}d^{[i]}$ of the $r$-th ground
state fulfill the TR transformation via
\begin{equation}
\bar{\omega}_{p}^{r|i|}\bar{d}^{[i]}=\mathcal{T}_{0,0}^{-1}\mathcal{Q}%
_{p/n}\left( \omega _{p}^{(n-r)|i|}d^{[i]}\right) \mathcal{Q}_{p/n}^{-1}%
\mathcal{T}_{0,0},
\end{equation}%
where the phases $\omega _{p}^{n|i|}$ have been canceled due to $\mathcal{Q}%
_{p/n}d^{[i]}\mathcal{Q}_{p/n}^{-1}=\omega _{p}^{-n|i|}d^{|i|}$. Since the
matrices $\omega _{p}^{(n-r)|i|}d^{[i]}$ in the right hand side generate the
$(n-r)$-th ground state, the TR symmetry transforms the $r$-th ground state
to the $(n-r)$-th ground state. Thus $\mathcal{T}_{r,r}=\mathcal{Q}%
_{p/n}^{-1}\mathcal{T}_{0,0}$ for $r=1,2,\cdots ,n-1$ and the permutation
matrix is $P_{ij}=\delta _{i,n-j}$.

Finally, we need to discuss the properties of $\mathcal{T}_{0,0}$, which
satisfies
\begin{equation}
\mathcal{T}_{0,0}\bar{\mathcal{T}}_{0,0}=(-1)^{\kappa }\mathbbm{1}
\end{equation}%
with $\kappa =0,1$. Since $\mathcal{T}_{0,0}$ also has a definite $\mathbb{Z}%
_{p/n}$ charge, we have
\begin{equation}
\mathcal{T}_{0,0}\mathcal{Q}_{p/n}=(-1)^{\mu }\mathcal{Q}_{p/n}\mathcal{T}%
_{0,0},
\end{equation}%
For odd $p/n$, the charge of $\mathcal{T}_{0,0}$ can only be zero, and there
are two different phases labelled by $\kappa =0,1$. While for even $p/n$,
the charge of $\mathcal{T}_{0,0}$ can be $0$ or $p/(2n)$, and there are four
phases labelled by $\kappa =0,1$ and $\mu =0,1$. Different from the previous
classification\cite{Meidan}, we find that the phases of $\mathbb{Z}_{p}$
parafermions with even $p$ do not always split into $4$ phases. For example,
the SSB phase of $\mathbb{Z}_{18}$ parafermions labelled by $n=6$ just
splits into two different gapped phases.

\section{Entanglement spectra of the TR enriched $\mathbb{Z}_{p}$
parafermionic phases}

It has been known that different topological phases can be featured by the
necessary degeneracies of the their ES\cite{Li&Haldane,ES}. Here we consider
the degeneracy of ES via a left-right bipartition of an infinite long chain,
so there is only one boundary in the reduced system. The topological
protected degeneracy of ES is determined by the structure and symmetries of
the reduced density matrix. According to the holographic principle, the
reduced density matrix obtained via the left-right bipartition of an
infinite long chain can be derived from the dominant eigenvectors of the
transfer operator $\mathbb{E}=\sum_{i}A^{[i]}\otimes \bar{A}^{[i]}$. We will
divide our discussion into three different cases. The corresponding results
have been summarized in Tab. \ref{GeneralCaseDeg}.

\subsection{For SPT phases}

\begin{figure}[tbp]
\centering
\includegraphics[width=8cm]{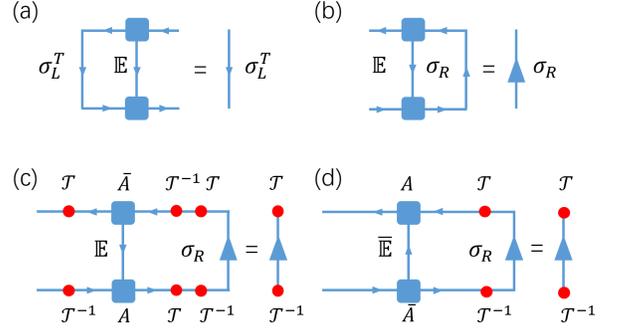}
\caption{The transfer operator and its left dominate eigenvector (a) and
right dominate eigenvector (b). (c) Applying the TR on the eigen-equation.
(d) Because $\mathcal{T}^{-1}\protect\sigma _{R}\mathcal{T}$ is the
eigenvector of $\bar{\mathbb{E}}$, $\mathcal{T}\protect\sigma _{R}^{T}%
\mathcal{T}^{-1}$ is also the eigenvector of $\mathbb{E}$.}
\label{TR-TM}
\end{figure}

These SPT phases are enriched from the trivial phase. According to the
properties of injective MPS\cite{MPS}, the dominant eigenvalue of $\mathbb{E}
$ is non-degenerate. We denote its left dominant eigenvector by $\sigma _{L}$
and the right dominant eigenvector by $\sigma _{R}$, as shown in Fig. 2(a)
and (b), respectively, and the dominant eigenvectors can be reshaped into
matrices. Because of the TR transformation satisfied by $A^{[i]}$ shown in
Eq.(\ref{TR}), the transfer operator $\mathbb{E}$ is TR symmetric, i.e.,
\begin{equation}
(\mathcal{T}\otimes \mathcal{T}^{-1})\bar{\mathbb{E}}(\mathcal{T}%
^{-1}\otimes \mathcal{T})=\mathbb{E},  \label{45}
\end{equation}%
where $\mathcal{T}$ effectively plays a role of TR transformation on virtual
degrees of freedom. Thus it takes the complex conjugation $\bar{\mathbb{E}}$
in the left hand side of Eq.(\ref{45}). By acting the TR on the
eigen-equations, we can demonstrate that $\mathcal{T}^{-1}\sigma _{L}^{T}%
\mathcal{T}$ and $\mathcal{T}\sigma _{R}^{T}\mathcal{T}^{-1}$ are also left
and right dominant eigenvectors of $\mathbb{E}$, as shown in Fig. \ref{TR-TM}
(c) and (d). Moreover, because $\sigma _{L}$ and $\sigma _{R}$ are Hermitian
operators, i.e., $\sigma _{L}^{T}=\bar{\sigma}_{L}$ and $\sigma _{R}^{T}=%
\bar{\sigma}_{R}$, and the dominant eigenvectors are unique, we have
\begin{equation}
\mathcal{T}^{-1}\bar{\sigma}_{L}\mathcal{T}=\sigma _{L},\mathcal{T}\bar{%
\sigma}_{R}\mathcal{T}^{-1}=\sigma _{R},
\end{equation}%
which manifest that $\sigma _{L}$ and $\sigma _{R}$ are also $\mathcal{T}$
symmetric.

Via an isometry map $U$, the entanglement Hamiltonian $H_{E}$ is given by $%
e^{H_{E}}=U^{\dagger }\sqrt{\bar{\sigma}_{L}}\sigma _{R}\sqrt{\bar{\sigma}%
_{L}}U$ for an infinite long chain\cite{Bultinck2017,ESofPEPS}. Since the
reduced density matrix $e^{H_{E}}$ and $\rho =\bar{\sigma}_{L}\sigma _{R}$
share the same eigenvalue spectrum, we study the spectrum properties $\rho $
for convenience. The $\mathcal{T\ }$symmetric $\sigma _{L}$ and $\sigma _{R}$
give rise to the $\mathcal{T}$ symmetric $\rho $:
\begin{equation}
\mathcal{T}\bar{\rho}\mathcal{T}^{-1}=\rho ,
\end{equation}%
where $\mathcal{T}$ is effectively anti-unitary. Then the different
behaviors of $\mathcal{T}$ give rise to the different topological protected
degeneracies.

We suppose that $v_{j}$ are the eigenstates of $\rho $ with eigenvalues $%
e_{j}$, i.e., $\rho v_{j}=e_{j}v_{j}$. Since $\rho $ is $\mathcal{T}$
symmetric, it can be easily derived that $\mathcal{T}\bar{v}_{j}$ are also
the eigenstates of $\rho $ with the same eigenvalues: $\rho \mathcal{T}\bar{v%
}_{j}=e_{j}\mathcal{T}\bar{v}_{j}$. Then we just need to determine whether $%
v_{j}$ and $\mathcal{T}\bar{v}_{j}$ describe the same states or not. Let us
first consider the effects of $\kappa $, which comes from $\mathcal{T}\bar{%
\mathcal{T}}=(-1)^{\kappa }\mathbbm{1}$. For $\kappa =0$, we need not to
distinguish the eigenstates $v_{j}$ and $\mathcal{T}\bar{v}_{j}$. However,
for $\kappa =1$, i.e., $\mathcal{T}\bar{\mathcal{T}}=-\mathbbm{1}$, $v_{j}$
and $\mathcal{T}\bar{v}_{j}$ are certainly different, guaranteed by the
Kramers theorem. So the eigenstates of $\rho $ form the Kramers pairs $%
\{(v_{j},\mathcal{T}\bar{v}_{j})\}$ and the ES is two-fold degenerate.

Then we consider the influences of $\mu $ defined by $\mathcal{T}\mathcal{Q}%
_{p}=(-1)^{\mu }\mathcal{Q}_{p}\mathcal{T}$. When $\mathcal{Q}_{p}$ and $%
\mathcal{T}$ commute, $\mathcal{T}$ doesn't change the charges of
eigenstates of $\rho $, namely $v_{j}$ and $\mathcal{T}\bar{v}_{j}$ have the
same charges. When $\mathcal{Q}_{p}$ and $\mathcal{T}$ anti-commute, the
charges of $v_{j}$ are shifted by $p/2$ under the action of $\mathcal{T}$.
Thus $v_{j}$ and $\mathcal{T}\bar{v}_{j}$ have different charges. They must
correspond to different states with the same eigenvalues $e_{j}$, and
two-fold degeneracy is produced in the ES.

Integrating the effects of both $\kappa $ and $\mu $, we conclude that as
long as one of $\kappa $ and $\mu $ is not zero, the ES must be at least
two-fold degenerate. In the situation that $\kappa =\mu =1$, the Kramers
pairs $\{(v_{j},\mathcal{T}\bar{v}_{j})\}$ consist of different charge
states, so the necessary degeneracy is not enlarged. So the ES is not
necessary degenerate for $\kappa =\mu =0$, otherwise it is at least two-fold
degenerate.

\subsection{For symmetry enriched topological phases}

Because the local matrices of the non-trivial topological phases have the
peculiar structures: $A^{i}=Y_{n}^{|i|}\otimes d^{i}$, we introduce the
sub-block transfer operator $\mathbb{E}^{\prime }=\sum_{i}d^{[i]}\otimes
\bar{d}^{[i]}$. The dominant left (right) eigenvector $\sigma _{L}^{\prime }$
($\sigma _{R}^{\prime }$) of $\mathbb{E}^{\prime }$ is non-degenerate,
because the MPS generated by $d^{[i]}$ are injective. Moreover, there are $n$%
-fold degenerate left and right dominant eigenvectors of $\mathbb{E}$, which
are given by $\sigma _{L,r}=Y_{n}^{r}\otimes \sigma _{L}^{\prime }$ and $%
\sigma _{R,r}=Y_{n}^{r}\otimes \sigma _{R}^{\prime }$ with $r\in \mathbb{Z}%
_{n}$. It can be proved that the spectrum of the whole reduced density
matrix is only determined by the charge-0 dominant eigenvectors\cite%
{Bultinck2017, XuZhang2}
\begin{equation}
\rho =\bar{\sigma}_{L,0}\sigma _{R,0}=\mathbbm{1}_{n}\otimes \bar{\sigma}%
_{L}^{\prime }\sigma _{R}^{\prime }=\mathbbm{1}_{n}\otimes \rho ^{\prime }.
\end{equation}%
So the ES has $n$-fold degeneracy without imposing the TR symmetry.

When discussing the degeneracy of ES with TR symmetry, we must know which
part of $\mu $ produces. For even $n$, $\mu $ comes from the non-trivial
part of the MPS. Considering the relation previously derived: $\mathcal{T}%
_{0}=\mathcal{Q}_{n}^{\mu }\otimes \mathcal{T}_{0,0}$, the sub-block
matrices $d^{[i]}$ obey the transformation $\bar{d}^{[i]}=\omega _{n}^{\mu
|i|}\mathcal{T}_{0,0}^{-1}d^{[i]}\mathcal{T}_{0,0}$, from which $\mathbb{E}%
^{\prime }$ is $\mathcal{T}_{0,0}$ symmetric:
\begin{equation}
\left( \mathcal{T}_{0,0}\otimes \mathcal{T}_{0,0}^{-1}\right) \bar{\mathbb{E}%
}^{\prime }\left( \mathcal{T}_{0,0}^{-1}\otimes \mathcal{T}_{0,0}\right) =%
\mathbb{E}^{\prime }.  \label{T00}
\end{equation}%
So $\mathcal{T}_{0,0}$ is irrelevant to the definition of $\mu $, the
symmetry of $\mathbb{E}^{\prime }$ as well as that of $\sigma _{L}^{\prime }$
and $\sigma _{R}^{\prime }$ have the same properties for $\mu =0,1$, and we
can predict that $\mu $ will not change the degeneracy of ES for even $n$.
Then the discussion for the spectrum of $\rho ^{\prime }$ is the same as
that of the last subsection, because $\rho ^{\prime }$ is also $\mathcal{T}%
_{0,0}$ symmetric:
\begin{equation}
\mathcal{T}_{0,0}\bar{\rho}^{\prime }\mathcal{T}_{0,0}^{-1}=\rho ^{\prime }.
\end{equation}%
Notice that $p/n$ must be odd for even $n$. Therefore, if $\kappa =0$, the
parafermion zero modes produce the $n$-fold degeneracy. If $\kappa =1$, the
eigenstates $v_{j}^{\prime }$ of $\rho ^{\prime }$ form the Kramers pairs $%
\{(v_{j}^{\prime },\mathcal{T}_{0,0}\bar{v}_{j}^{\prime })\}$, so the total
degeneracy of ES is $2n$.

For the odd $n$ case, $\mu $ comes from the trivial part and is defined by $%
\mathcal{T}_{0,p/n}\mathcal{Q}_{p/n}=(-1)^{\mu }\mathcal{Q}_{p/n}\mathcal{T}%
_{0,p/n}$, and Eq.(\ref{T00}) is still satisfied. Because $\mu $ must be $0$
for odd $p/n$. the ES is $n$-fold degenerate for $\kappa =0$ and $2n$-fold
degenerate for $\kappa =1$. For even $p/n$, the $n$-fold degenerate ES is
contributed by the parafermion zero edge modes for $\kappa =0$ and $\mu =0$,
otherwise there is a $2n$-fold degenerate ES protected by both parafermion
zero edge modes and SPT order.

\subsection{For SSB coexisting SPT phases}

In this case, all sub-blocks have the same sub-block transfer operator $%
\mathbb{E}^{\prime }=\sum_{i}d^{[i]}\otimes \bar{d}^{[i]}$, and the reduced
density matrix for whole ground state subspace is thus a direct sum of the
reduced density matrices for individual ground states,
\begin{equation}
\rho =\bigoplus\limits_{r=0}^{n-1}\rho ^{\prime
}=\bigoplus\limits_{r=0}^{n-1}\bar{\sigma}_{L}^{{\prime }}\sigma
_{R}^{\prime },  \label{SSB-rho}
\end{equation}%
where $\sigma _{L}^{\prime }$ and $\sigma _{R}^{\prime }$ are the left and
right dominant eigenvectors of the sub-block transfer operator $\mathbb{E}%
^{\prime }$. Because the system just picks one of the degenerate ground
states, we just consider the ES of $\rho ^{\prime }$. The corresponding
analysis is the same as that of SPT phases. So there is no topological
degeneracy if $\kappa =$ $\mu =0$, otherwise the ES has $2$-fold degeneracy.
$\kappa $ and $\mu $ have the same interpretations as the SPT phases.

\section{Discussion and Conclusion}

Actually, for the SPT phases and the SSB coexisting SPT phases, the MPS have
the same properties as those of bosonic MPS in one dimension. So their
classification can be fitted into the framework of second group cohomology
classifying bosonic SPT phases\cite{Chen-Gu-Liu-Wen(2013)}. In fact, the SPT
phases enriched from the trivial phase are classified by the second group
cohomology $H^{2}(\mathbb{Z}_{p}\rtimes \mathbb{Z}_{2}^{T},U(1))=\mathbb{Z}%
_{2}\times \mathbb{Z}_{\text{gcd}(p,2)}$, where $\mathbb{Z}_{2}^{T}$ is the
TR symmetry group and \textquotedblleft gcd\textquotedblright\ denotes the
greatest common divisor. The SSB coexisting SPT phases are also classified
by the subgroup $\mathbb{Z}_{p/n}\rtimes \mathbb{Z}_{2}^{T}$ and SPT order $%
H^{2}(\mathbb{Z}_{p/n}\rtimes \mathbb{Z}_{2}^{T},U(1))=\mathbb{Z}_{2}\times
\mathbb{Z}_{\text{gcd}(p/n,2)}$ under this subgroup.

However, the classification of those symmetry enriched topological phases is
different from those bosonic MPS. Employing the recent classification for
the one-dimensional interacting fermions with on-site symmetries\cite%
{Bultinck2017,Kapustin2017}, we can also integrate our results for the
one-dimensional parafermion systems into this generalized framework.
Therefore, the $\mathbb{Z}_{p}$ symmetric non-trivial topological phases
labelled by $n$ with the TR symmetry can be classified by both $H^{2}(%
\mathbb{Z}_{p/n}\rtimes \mathbb{Z}_{2}^{T},U(1))=\mathbb{Z}_{2}\times
\mathbb{Z}_{\text{gcd}(p/n,2)}$ and $H^{1}(\mathbb{Z}_{2}^{T},\mathbb{Z}%
_{n})=\mathbb{Z}_{\text{gcd}(n,2)}$, where $\mathbb{Z}_{2}$, $\mathbb{Z}_{%
\text{gcd}(p/n,2)}$ and $\mathbb{Z}_{\text{gcd}(n,2)}$ correspond to the
indices $\kappa $, $\mu $ and $\mu ^{\prime }$ defined in Section V.B,
respectively. Here the second group cohomology $H^{2}(\mathbb{Z}%
_{p/n}\rtimes \mathbb{Z}_{2}^{T},U(1))$ labels the SPT order under the
symmetry group $\mathbb{Z}_{p/n}\rtimes \mathbb{Z}_{2}^{T}$, and the first
group cohomology $H^{1}(\mathbb{Z}_{2}^{T},\mathbb{Z}_{n})$ just classifies
the actions of TR symmetry on the edge modes. Moreover, the non-trivial SPT
order given by the second group cohomology double the degeneracy of the ES.

In conclusion, using the parafermionic MPS, we have established the complete
classification of one-dimensional gapped phases of $\mathbb{Z}_{p}$
parafermions \textit{with} the TR symmetry satisfying $T^{2}=1$. The
possible gapped phases are enriched and classified by three indices $n$, $%
\kappa $ and $\mu $, where $n$ is a divisor of $p$, $\kappa \in \mathbb{Z}%
_{2}$ corresponds to the linear or projective actions of TR symmetry on the
edges, and $\mu \in \mathbb{Z}_{2}$ describes the commutation relations
between the TR and (fractionalized) charge operator at the virtual degrees
of freedom. For the $\mathbb{Z}_{r}$ symmetric ground states, where $r=p$
for trivial or topological phase, and $r=p/n$ for SSB phases, the original
gapped phases with odd $r$ are divided into two phases, while those phases
with even $r$ are separated into four phases. The resulting gapped
parafermionic phases include the SPT phases, symmetry enriched topological
phases, and the SSB coexisting SPT phases. How to realize these novel phases
in physical systems will be our future research investigations.

\textit{Acknowledgment.- }The authors would like to thank Guo-Yi Zhu and
Zi-Qi Wang for their stimulating discussion and acknowledges the support of
National Key Research and Development Program of China (No.2017YFA0302902).

\end{document}